\documentclass[aps,pra,twocolumn,superscriptaddress]{revtex4}
\usepackage{amssymb}
\usepackage{amsmath}
\usepackage{graphicx}
\usepackage{epsfig}
\usepackage{subfigure}
\usepackage{xcolor}

\begin{document}

\title{$\mathcal{PT}$ symmetric phase transition and single-photon transmission in an optical trimmer system}
\author{L. F. {Xue}}
\affiliation{Center for Quantum Sciences and School of Physics and Center for Advanced Optoelectronic Functional Materials Research and Key Laboratory for UV Light-Emitting Materials and Technology of Ministry of Education, Northeast Normal University, Changchun 130024, China}
\author{Z. R. Gong}
\affiliation{College of Physics and Energy, Shenzhen University, Shenzhen 518060, China}
\author{H. B. {Zhu}}
\email{hbzhu@nenu.edu.cn}
\affiliation{Center for Quantum Sciences and School of Physics and Center for Advanced Optoelectronic Functional Materials Research and Key Laboratory for UV Light-Emitting Materials and Technology of Ministry of Education, Northeast Normal University, Changchun 130024, China}
\author{Z. H. {Wang}}
\email{wangzh761@nenu.edu.cn}
\affiliation{Center for Quantum Sciences and School of Physics and Center for Advanced Optoelectronic Functional Materials Research and Key Laboratory for UV Light-Emitting Materials and Technology of Ministry of Education, Northeast Normal University, Changchun 130024, China}

\begin{abstract}
The parity-time ($\mathcal{PT}$) symmetric structures have exhibited potential applications in developing various robust quantum devices. In an optical trimmer with balanced loss and gain, we analytically study the $\mathcal{PT}$ symmetric phase transition by investigating the spontaneous symmetric breaking. We also illustrate the single-photon transmission behaviors in both of the $\mathcal{PT}$ symmetric and  $\mathcal{PT}$ symmetry broken phases. We find (i) the non-periodical dynamics of single-photon transmission in the $\mathcal{PT}$ symmetry broken phase instead of $\mathcal{PT}$ symmetric phase can be regarded as a signature of phase transition; and (ii) it shows unidirectional single-photon transmission behavior in both of the phases but comes from different underlying physical mechanisms. The obtained results may be useful to implement the photonic devices based on coupled-cavity system.
\end{abstract}

\maketitle

\section{Introduction}

Since Bender and Boettcher proposed the concept of parity-time ($\mathcal{PT}$) symmetry~\cite{bender1,bender2}, it has attracted a lot of attentions due to its potential applications. Remarkably, the system with $\mathcal{PT}$ symmetry can undergo a phase transition when the parameter that controls the non-Hermiticity surpasses a critical value which is usually called exceptional point (EP)~\cite{bender2,AM}. Below the EP, all of the eigen-values of the non-Hermit Hamiltonian are real, and some or all of the eigen values become complex beyond the EP.

The broken of $\mathcal{PT}$ symmetry will lead to a lot of interesting phenomena. For example,  people have observed the non-reciprocal photonic transmission in the $\mathcal{PT}$ symmetric structure~\cite{long,xiaomin,wujh,songzhi,HR,ZL,LF1,NB1}, and predicted the enhancement of nonlinear interaction due to field localization~\cite{jiahua1,jiahua2}.  Moreover, the unique property of the system with $\mathcal{PT}$ symmetry has potential application in various fields, such as loss-induced or gain-induced transparency~\cite{Guo,Jing}, efficient photon or phonon lasing~\cite{BP,LF,HH,HJ,BH}, ultralow-threshold optical chaos~\cite{CT, xinyou} as well as quantum metrology~\cite{jingzhang}.

On the other hand, the coupled-cavity system is widely used to coherently control the photon transfer. In the coupled-cavity-array with infinite length, the defect can be introduced to construct a singe-photon switcher~\cite{sun1,wang0}, router~\cite{sun2} and frequency converter~\cite{wang}. Moreover, within the capacity of current experiments, a lot of attentions have been paid on the optical dimmer (also called optical molecule~\cite{MB,YP}, which is composed of two coupled cavities), such as the coherent polariton~\cite{xiao} and state transfer~\cite{state}. Furthermore, motivated by the simulation of photosynthesis harvest system
in recent years, there are also studies about the photon~\cite{ultrastrong} and thermal transport~\cite{thermal} in the trimmer structure.

In this paper, we focus on an optical trimmer with $\mathcal{PT}$ symmetry. Here, our scheme is composed of an active gain cavity and a passive loss cavity, which simultaneously couple with a third cavity without loss and gain, to form a coupled-cavity-array as shown in Fig.~\ref{trimmer}. With balanced gain and loss, which are described phenomenally in this paper, we show a $\mathcal{PT}$ symmetric phase transition in the non-Hermitian system.  In the $\mathcal{PT}$ symmetric phase, all of the eigen-values of the non-Hermitian Hamiltonian are real and the corresponding eigen-states show the same $\mathcal{PT}$ symmetric character as that of the Hamiltonian. On the contrary, in the $\mathcal{PT}$ symmetry broken phase, the complex eigen-values emerge, and the $\mathcal{PT}$ symmetry of the eigen-states disappear. In this sense, the system will undergo a spontaneous symmetric breaking as the phase transition occurs.  In addition, the $\mathcal{PT}$ symmetric phase transition is accompanied by the field localization. That is, the photon shows an equal weight distribution between the passive and active cavity in $\mathcal{PT}$ symmetric phase and localized at the active cavity in $\mathcal{PT}$ symmetry broken phase. Compared with the phase transition in optical molecule or dimmer~\cite{long,xiaomin,jiahua1,jiahua2}, the critical coupling strength in optical trimmer is much smaller, and is therefore easier for experimental realization.

In such a system, we study the unidirectional single-photon transmission in both of the $\mathcal{PT}$ symmetric and $\mathcal{PT}$ symmetry broken phases. In the $\mathcal{PT}$ symmetric phase, it shows a periodical oscillation, and the unidirectional
transmission results from the  breaking of time reversal symmetry. In the
$\mathcal{PT}$ broken symmetric phase, where the gain compensates the loss~\cite{long}, the single-photon transfer exhibits a non-periodical feature and the unidirectional transmission stems mainly from the field localization.

The rest of the paper is organized as follows. In Sec.~\ref{model}, we present a $\mathcal{PT}$ symmetry model by an optical trimmer with balanced loss and gain and discuss the
$\mathcal{PT}$ symmetric phase transition. In Sec.~\ref{transmission}, we illustrate the
unidirectional single-photon transmission in both of the $\mathcal{PT}$ symmetric phase and $\mathcal{PT}$ symmetry broken phase. At last, we give a brief conclusion in Sec.~\ref{conclusion}.

\section{Model and $\mathcal{PT}$ phase transition}
\label{model}
\begin{figure}[tbp]
\begin{centering}
\includegraphics[width=8cm]{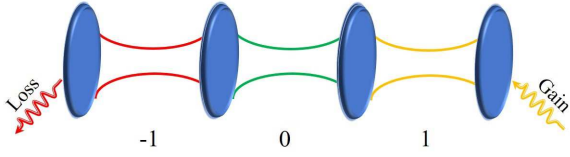}
\end{centering}
\caption{(Color online) Schematic illustration of $\mathcal{PT}$ symmetric optical trimmer. In this setup, the passive cavity $-1$ and active cavity $1$ simultaneously couple to the central cavity $0$, which is without loss or/and gain.}
\label{trimmer}
\end{figure}

As shown in Fig.~\ref{trimmer}, our model consists of an array of three single-mode cavities~\cite{ultrastrong}, where a passive and an active
cavity (labelled by ``$-1$" and ``$1$" respectively) simultaneously couple to the third cavity (labelled by ``$0$") without loss and gain. By describing the gain and loss in our scheme
phenomenologically, the Hamiltonian is written as
\begin{eqnarray}
H&=&(\omega_{-1}-i\gamma_{-1})a_{-1}^{\dagger}a_{-1}+\omega_{0}a_{0}^{\dagger}a_{0}
+(\omega_{1}+i\gamma_1)a_{1}^{\dagger}a_{1}\nonumber\\&&+
J(a_{-1}^{\dagger}a_0+a_0^{\dagger}a_{1}+a_{0}^{\dagger}a_{-1}+a_1^{\dagger}a_{0}),
\label{PTH}
\end{eqnarray}
where $a_l (l=-1,0,1)$ is the annihilation operator for the $l$th cavity with resonant frequency
$\omega_l$. $J$ is the photon-tunneling strength between the two nearest cavities, which can be
adjusted by changing the distance of them. In addition, we use $\gamma_{-1}(>0)$ to denote the decay of the passive cavity and $\gamma_1(>0)$ to denote the gain of active cavity. Hereafter, we consider the
case that $\omega_{-1}=\omega_0=\omega_1=\omega$ and $\gamma_{-1}=\gamma_{1}=\gamma$, therefore the Hamiltonian
satisfies a $\mathcal{PT}$ symmetry, that is $[H, \mathcal{PT}]=0$. Here $\mathcal{P}$ represents the mirror
reflection $1\leftrightarrow-1$ and $\mathcal{T}$ denotes the time reversal $i\leftrightarrow-i$~\cite{bender2}.

To deeply investigate the $\mathcal{PT}$ phase transition in our system, we write the Hamiltonian in the form of
\begin{equation}
H=\left(\begin{array}{ccc}
a_{1}^{\dagger} & a_{2}^{\dagger} & a_{3}^{\dagger}\end{array}\right)\mathcal{H}\left(\begin{array}{c}
a_{1}\\
a_{2}\\
a_{3}
\end{array}\right),
\end{equation}
where
\begin{equation}
\mathcal{H}=\left(\begin{array}{ccc}
\omega-i\gamma & J & 0\\
J & \omega & J\\
0 & J & \omega+i\gamma
\end{array}\right).
\label{mh}
\end{equation}
Solving the secular equation ${\rm det}(\mathcal{H}-EI)=0$, where $I$ is a $3\times3$ identity
matric, we will obtain the eigenvalues and the corresponding eigenstates of the Hamiltonian $\mathcal{H}$,
yielding
\begin{eqnarray}
E_0&=&\omega,\,\,|E_0\rangle=\frac{1}{\sqrt{2+(\gamma/J)^2}}(-1,-i\frac{\gamma}{J},1),\\
E_{\pm}&=&\omega\pm\sqrt{2J^2-\gamma^2},\,\,|E_{\pm}\rangle=\frac{1}{\mathcal{N_{\pm}}}(a_{\pm},b_{\pm},1).
\end{eqnarray}
where $\mathcal{N}_0$ and $\mathcal{N}_{\pm}$ are the normalized constants and we have defined
\begin{eqnarray}
a_{\pm}&:=&\frac{J^2-\gamma^2\mp i\gamma\sqrt{2J^2-\gamma^2}}{J^2},\\
b_{\pm}&:=&\frac{-i\gamma\pm\sqrt{2J^2-\gamma^2}}{J}.
\end{eqnarray}

We note that  the eigenvalue $E_0$ is always real and is independent of $J$ and $\gamma$, and
the eigen-state satisfies the $\mathcal{PT}$ symmetry, that is $\mathcal{PT}|E_0\rangle=e^{i\pi}|E_0\rangle$.  However, the other paired eigenvalues $E_{\pm}$ are dependent not only on $\omega$, but also on $\gamma$ and $J$.

To obtain a purely real spectrum, we need a strong inter-cavity coupling strength, i.e., $J>\gamma/\sqrt{2}$. It then satisfies $\mathcal{N}_+=\mathcal{N}_-=2$ and the eigen-state $|E_\pm\rangle$ can be simplified as
\begin{equation}
|E_\pm\rangle=\frac{1}{2}(e^{\mp i\theta_1},2e^{\mp i\theta_2},1),
\end{equation}
where $\theta_1:=\arctan[\gamma\sqrt{2J^2-\gamma^2}/(J^2-\gamma^2)]$ and $\theta_2:=\arctan[\sqrt{2J^2-\gamma^2}/\gamma]$.
A simple calculation tells us
\begin{equation}
\mathcal{PT}|E_\pm\rangle=e^{\pm i\theta_1}|E_\pm\rangle,
\end{equation}
which implies that the wave functions $|E_\pm\rangle$ are transformation invariant under
the $\mathcal{PT}$ operation (except for a global phase).
On the other hand, for the situation of $J<\gamma/\sqrt{2}$, the imaginary parts of $E_{\pm}$ emerge and $E_{\pm}=\omega\pm i\sqrt{\gamma^2 -2J^2}$. Meanwhile, we can not find a global phase $\phi$ to satisfy $\mathcal{PT}|E_\pm\rangle=e^{\pm i\phi}|E_\pm\rangle$. In other words, when the system undergoes a spontaneous symmetry breaking as the inter-cavity coupling crosses the EP $J=\gamma/\sqrt{2}$. In this sense, we name the phase in the regime $J>\gamma/\sqrt{2}$ as the $\mathcal{PT}$ symmetric phase and that for $J<\gamma/\sqrt{2}$ as the $\mathcal{PT}$ symmetry broken phase.

\begin{figure}[tbp]
\centering
\includegraphics[width=8cm]{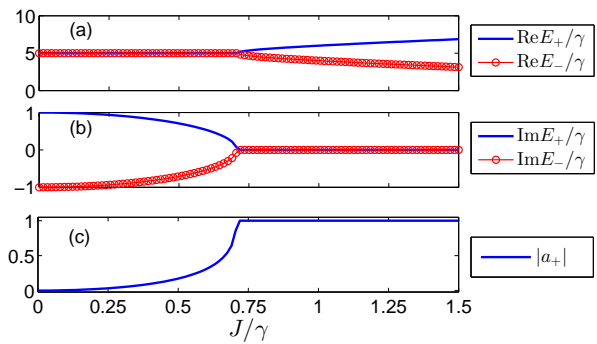}
\caption{(Color online) The real parts of $E_{\pm}$ (a), imaginary parts of $E_{\pm}$ (b),
and $|a_+|$ as a function of the coupling strength $J$. We have chosen
$\omega=5\gamma$, and all of the other parameters are in units of $\gamma=1$.}
\label{eigs}
\end{figure}

In Figs.~\ref{eigs} (a) and (b) , we give the real and imaginary parts of $E_{\pm}$ respectively. In the $\mathcal{PT}$ symmetry broken phase ($J<\gamma/\sqrt{2}$), the small coupling strength protects the gained energy flowing from the active cavity to the passive one, and the long lifetime supermode $E_+$ is localized at the active cavity as shown in Fig.~\ref{eigs} (c) , where we plot $|a_+|$ as a function of the coupling strength $J$. On contrary, in the $\mathcal{PT}$ symmetric phase ($J>\gamma/\sqrt{2}$) , the gained energy is transferred to the passive cavity quickly, and the photon yields an equal weight distribution in the passive and active cavities, that is $|a_\pm|\equiv1$.

We point out that here, the similar $\mathcal{PT}$ symmetric phase transition also occurrs in optical dimmer which consists of two cavities with balanced loss and gain (see Refs.~\cite{long,xiaomin} and the references therein).  The differences stems in the following two aspects: On the one hand, in our optical trimmer system, there exist a single real energy level $E_0$, and the corresponding wave function is always invariant under the $\mathcal{PT}$ operation, independent of whether the phase transition occurs. On the other hand, the EP of optical trimmer system is $J=\gamma/\sqrt{2}$ instead of $J=\gamma$ in optical dimmer~\cite{long,xiaomin,jiahua1,jiahua2}, that is, a smaller coupling strength is needed, which is more easily to be realized in experiments.

\section{single-photon transmission}
\label{transmission}

To show the effect of $\mathcal{PT}$ symmetric phase transition on the dynamics of the system, we in this section consider the different behaviors of single-photon transmission when it is excited in the passive or active cavity initially, in both of the $\mathcal{PT}$ symmetric and $\mathcal{PT}$ symmetry broken phases.

\begin{figure}[tbp]
\centering
\includegraphics[width=8cm]{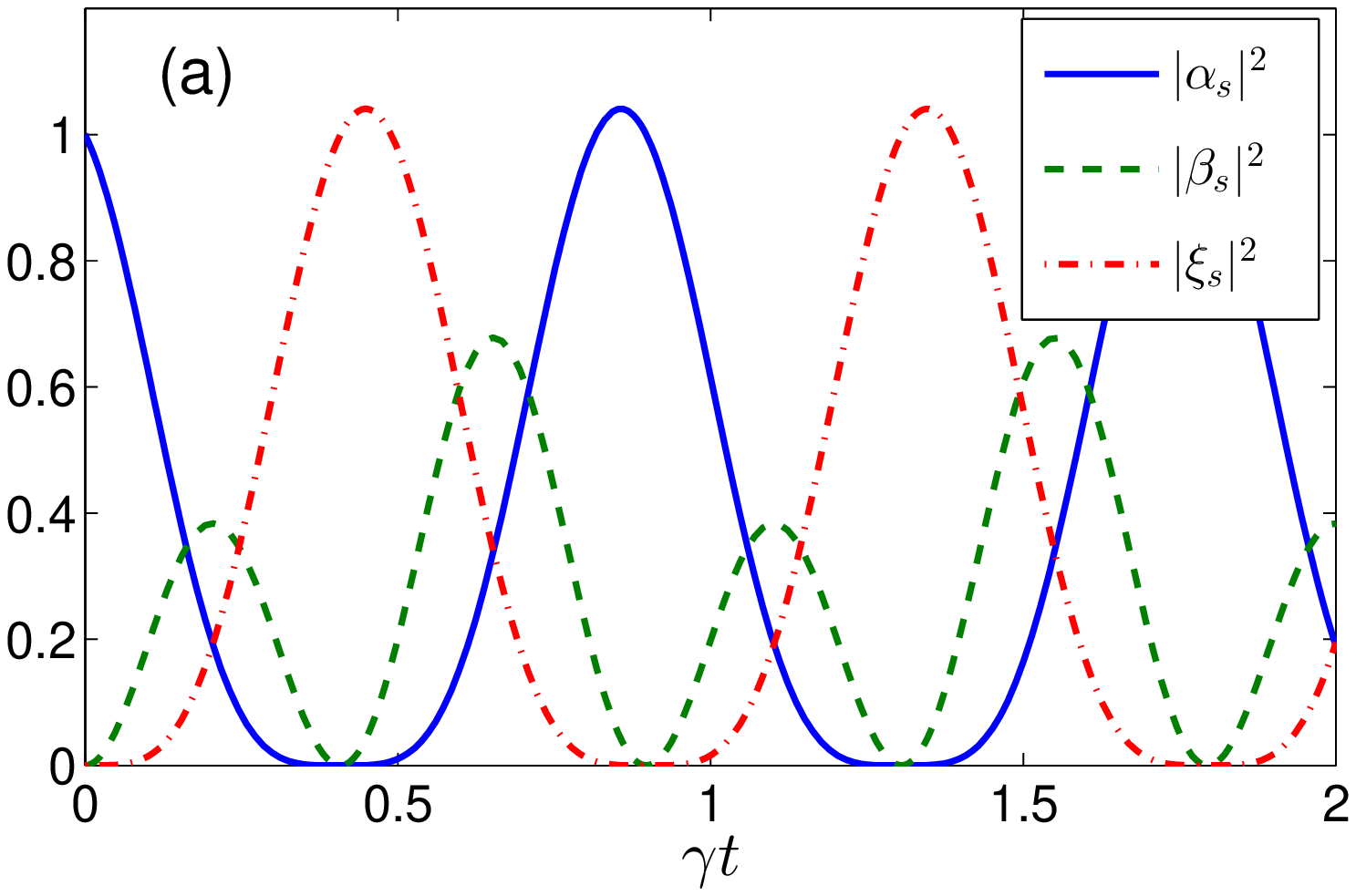}
\includegraphics[width=8cm]{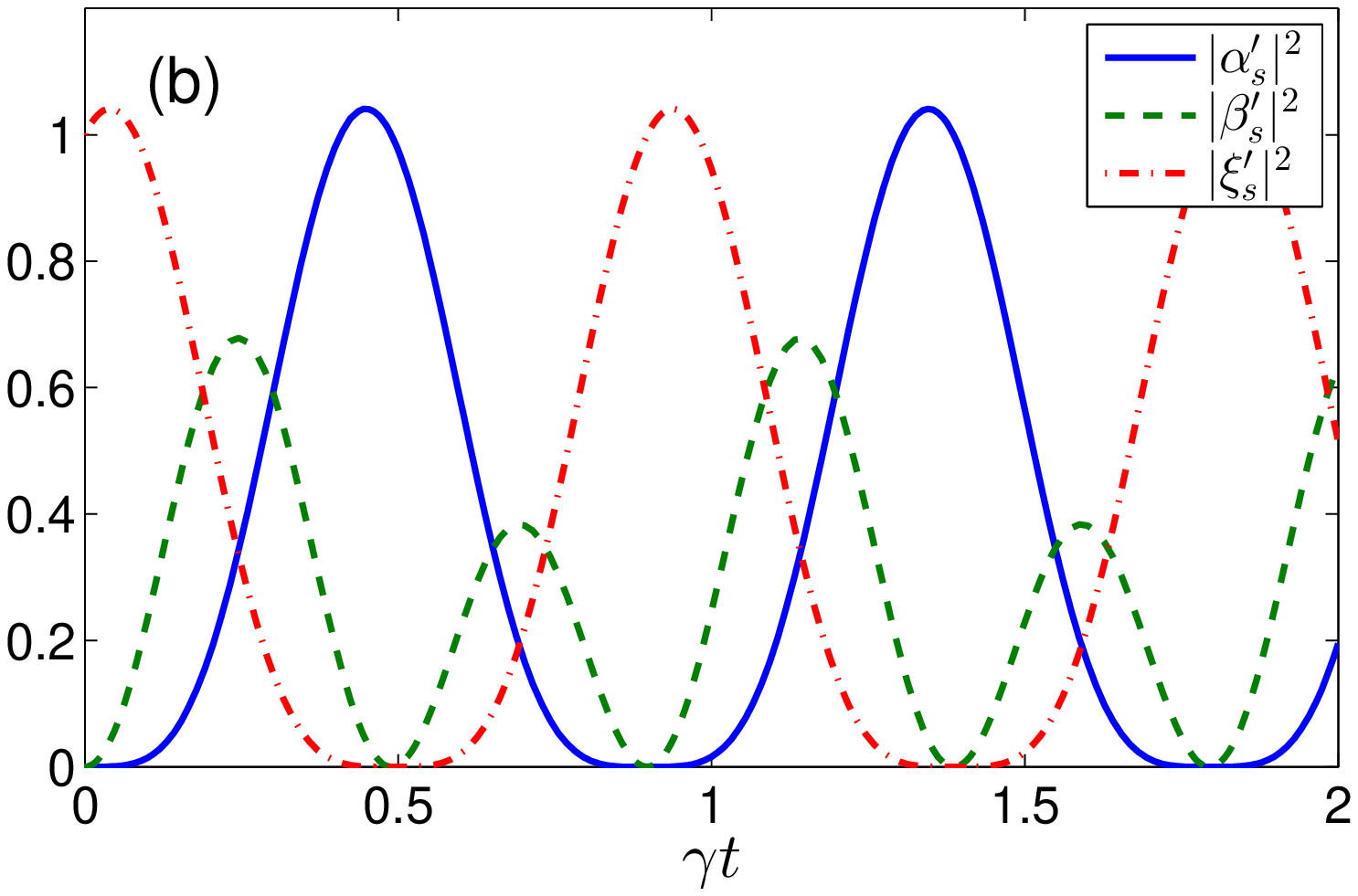}
\caption{(Color online) The illustration of dynamics of the system when the single photon
is initially excited in the passive (a) and active (b) cavities. The parameters are set as
$\omega=5\gamma, J=5\gamma$, and all parameters are in units of $\gamma=1$. Under these parameters, the system is in the $\mathcal{PT}$ symmetric phase.}
\label{ss}
\end{figure}

Since we only consider the transmission of single photon in this paper, the wave function at arbitrary time $t$
can be assumed as
\begin{equation}
|\psi(t)\rangle=\alpha(t)|1;0;0\rangle+\beta(t)|0;1;0\rangle+\xi(t)|0;0;1\rangle,
\end{equation}
where $|m;n;q\rangle:=|m\rangle_{-1}\otimes|n\rangle_{0}\otimes|q\rangle_{1}$, and $|n\rangle_{i}(i=-1,0,1)$ represents that the $i$th cavity is in the Fock state $|n\rangle$.

Governed by the Hamiltonian in Eq.~(\ref{PTH}), the dynamics of the system is determined by the
Schoedinger equation $i\partial_t|\psi\rangle=H|\psi\rangle$, which gives
\begin{subequations}
\begin{eqnarray}
i\frac{d}{dt}\alpha(t)&=&(\omega-i\gamma)\alpha(t)+J\beta(t),\\
i\frac{d}{dt}\beta(t)&=&\omega\beta(t)+J[\alpha(t)+\xi(t)],\\
i\frac{d}{dt}\xi(t)&=&(\omega+i\gamma)\xi(t)+J\beta(t).
\end{eqnarray}
\label{cof}
\end{subequations}

\subsection{Single-photon transmission in $\mathcal{PT}$ symmetric phase}
We now study the behavior of the single-photon transmission in the $\mathcal{PT}$ symmetric phase, that is $J>\gamma/\sqrt{2}$. Firstly, we consider the situation that the photon is excited in the passive cavity initially, that is $\alpha(0)=1,\beta(0)=\xi(0)=0$, we can obtain explicitly the probability amplitudes for finding a photon  in the three cavities as
\begin{eqnarray}
\alpha_s(t)&=&\frac{2J^2e^{-i\omega t}}{\Delta^2}\cos^2(\frac{\Delta t+\phi_1}{2}),\nonumber \\ \\
\beta_s(t)&=&\frac{2 iJ^2e^{-i\omega t}}{\Delta^2}\sin(\frac{\Delta t}{2})\sin(\frac{\Delta t-\phi_2}{2}),\\
\xi_s(t)&=&-\frac{2J^2e^{-i\omega t}}{\Delta^2}\sin^2(\frac{\Delta t}{2}),
\end{eqnarray}
where $\Delta=\sqrt{2J^2-\gamma^2},\phi_1=\arctan[\Delta \gamma/(J^2-\gamma^2)],\phi_2=2\arctan(\Delta/\gamma)$.
Secondly, when the single photon is initially excited in the active cavity, that is $\alpha(0)=\beta(0)=0,\xi(0)=1$, the solution of Eqs.~(\ref{cof}) are obtained as
\begin{eqnarray}
\alpha'_s(t)&=&-\frac{2J^2e^{-i\omega t}}{\Delta^2}\sin^2(\frac{\Delta t}{2})
, \\
\beta'_s(t)&=&\frac{2 iJ^2e^{-i\omega t}}{\Delta^2}\sin(\frac{\Delta t}{2})\sin(\frac{\Delta t+\phi_2}{2}),\\
\xi'_s(t)&=&
\frac{2J^2e^{-i\omega t}}{\Delta^2}\cos^2(\frac{\Delta t-\phi_1}{2}).
\end{eqnarray}

In Figs.~\ref{ss} (a) and (b), we plot the corresponding probabilities $|h_s(t)|^2$ and $|h'_s(t)|^2$ $(h=\alpha,\beta,\xi)$ as functions of evolution time $t$. As shown in
the figure, when the system is in the $\mathcal{PT}$ symmetric phase, the dynamics shows regular periodical oscillations. If the single photon is excited in the passive cavity, i.e., $\alpha(0)=1$, the decay makes $|\alpha_s(t)|^2$ directly decrease to zero and then the revival occurs. However, if it is excited in the active cavity, i.e., $\xi(0)=1$, with the assistance of the gain effect, $|\xi'_s(t)|^2$ firstly reaches its maximal value $[2J^2/(2J^2-\gamma^2)]^2$, which is obviously larger than $1$ and then oscillates between the maximal value and zero. As for the central cavity, we can also observe that $\beta_s(t)\neq\beta'_s(t)$. In this sense, our system exhibits a unidirectional phenomenon in the single-photon level even in the $\mathcal{PT}$ symmetric phase. Furthermore, we find that the initial phases $\phi_1$ and $\phi_2$ are $\gamma$ dependent and $\phi_1(-\gamma)=-\phi_1(\gamma),\phi_2(-\gamma)=-\phi_2(\gamma)$. As a result, by regarding the  amplitudes as functions of $t$ and $\gamma$, we will reach the relationship
$\alpha_s(t,-\gamma)=\xi'_s(t,\gamma),\beta_s(t,-\gamma)=\beta'_s(t,\gamma)$ and $\xi_s(t,-\gamma)=\alpha'_s(t,\gamma)$.  Meanwhile, it is obvious from Eq.~(\ref{mh}) that $\mathcal{H}(i\leftrightarrow-i)=\mathcal{H}(\gamma\leftrightarrow-\gamma)$, therefore, the
single-photon unidirectional transmission in $\mathcal{PT}$ symmetric phase comes from the breaking of time reversal symmetry, that is $[\mathcal{T},\mathcal{H}]\neq0$.

\subsection{Single-photon transmission in $\mathcal{PT}$ symmetry broken phase}

In this subsection, we will continue to study the dynamics of single-photon transmission in the $\mathcal{PT}$ symmetry broken phase, where $J<\gamma/\sqrt{2}$. On the one hand, we consider that the single photon is initially excited in the passive cavity, then the dynamics of system is obtained as
\begin{eqnarray}
\alpha_b(t)&=&-\frac{e^{-i\omega t}}{\delta^2}[J^2+(J^2-\gamma^2)\cosh(\delta t)+\gamma\delta\sinh(\delta t)],\nonumber \\ \\
\beta_b(t)&=&-\frac{iJe^{-i\omega t}}{\delta^2}[\gamma\cosh(\delta t)-\gamma-\delta\sinh(\delta t)],\\
\xi_b(t)&=&-\frac{2J^2e^{-i\omega t}\sinh^2(\frac{\delta t}{2})}{\delta^2}.
\end{eqnarray}
where $\delta=\sqrt{\gamma^2-2J^2}$. On the other hand, when the single photon is initially excited in the active cavity, the dynamics of the system is described by
\begin{eqnarray}
\alpha'_b(t)&=&-\frac{2J^2e^{-i\omega t}\sinh^2(\frac{\delta t}{2})}{\delta^2},\\
\beta'_b(t)&=&-\frac{iJe^{-i\omega t}}{\delta^2}[\gamma\cosh(\delta t)-\gamma+\delta\sinh(\delta t)],\\
\xi'_b(t)&=&-\frac{e^{-i\omega t}}{\delta^2}[J^2+(J^2-\gamma^2)\cosh(\delta t)-\gamma\delta\sinh(\delta t)].\nonumber \\
\end{eqnarray}

\begin{figure}[tbp]
\centering
\includegraphics[width=8cm]{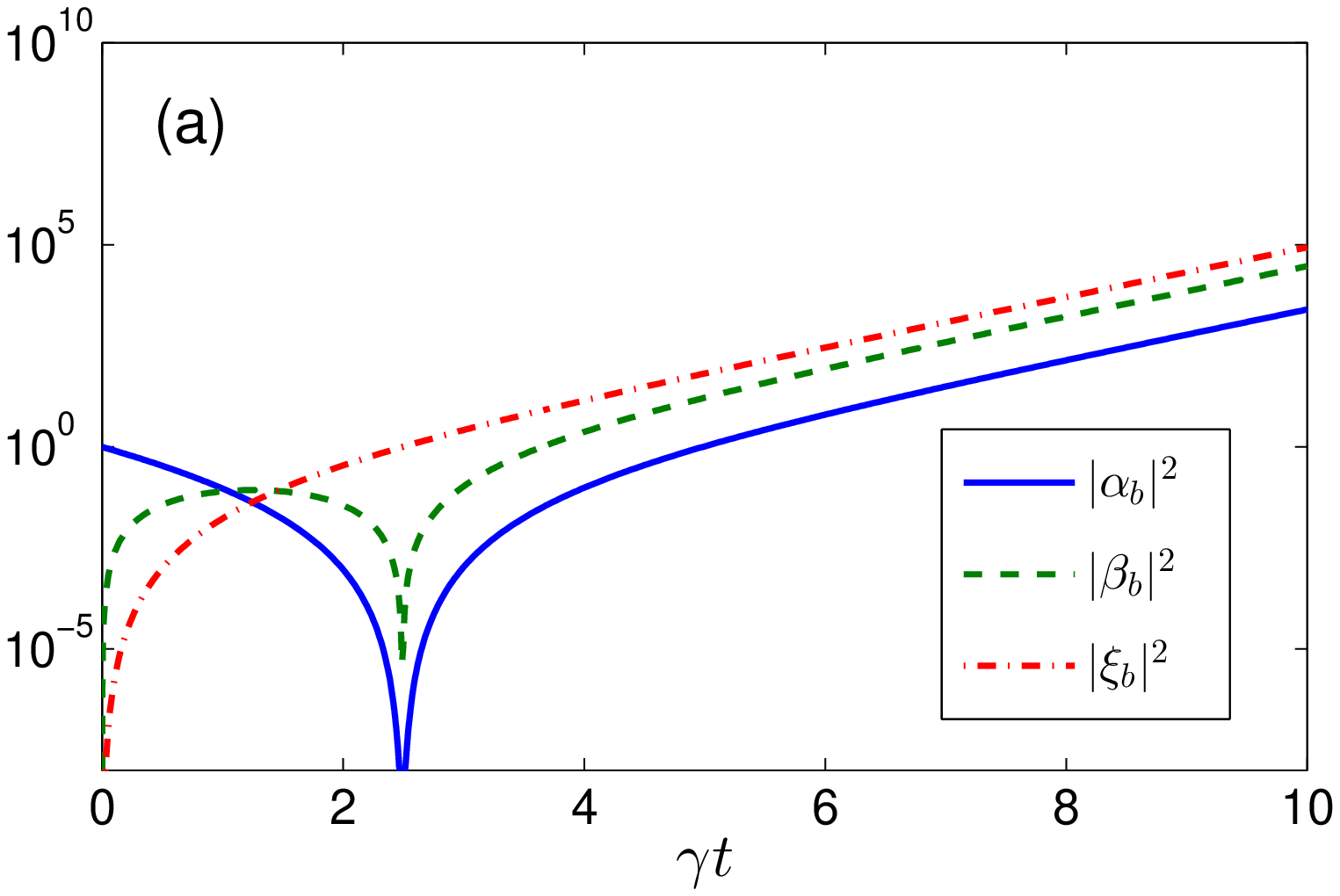}
\includegraphics[width=8cm]{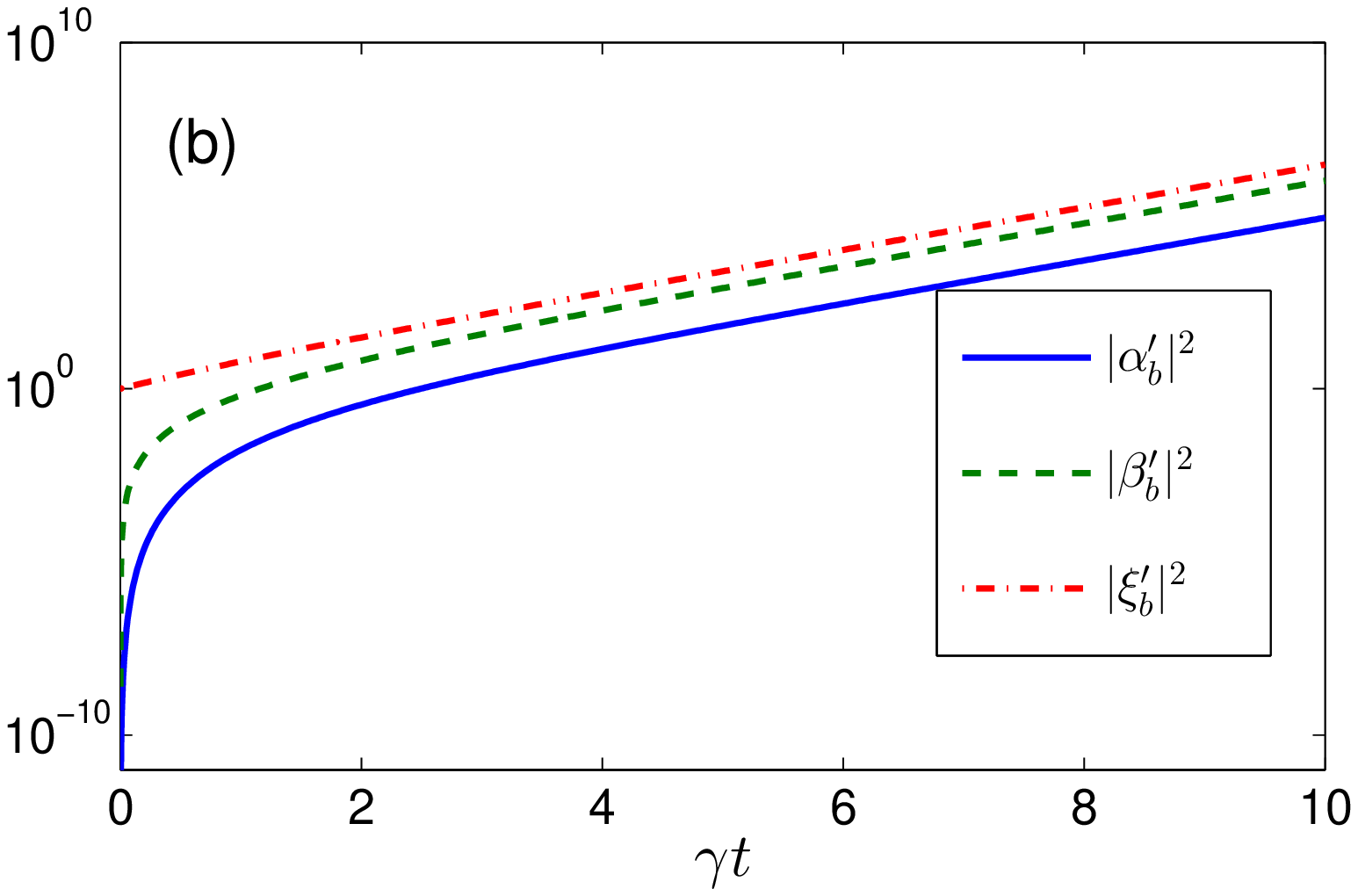}
\caption{(Color online) The illustration of dynamics of the system when the single photon
is initially excited in the passive (a) and active (b) cavities. The parameters are set as
$\omega=5\gamma, J=0.5\gamma$, and all parameters are in units of $\gamma=1$. Under these parameters, the system is in the $\mathcal{PT}$ symmetry broken phase.}
\label{bb}
\end{figure}
 In Fig.~\ref{bb}, we depict the dynamics of the system when it is in the $\mathcal{PT}$ symmetry broken phase. Obviously, it shows a completely different behavior compared with the case in the $\mathcal{PT}$ symmetric phase.  As shown in Fig.~\ref{bb}(a), when the single photon is initially excited in the passive cavity, the photon will experience a loss firstly and then the gain in the active cavity will compensate the loss. As a result, the probability for finding the photon in the cavities will  increase as the time elapse. As for the central cavity, the incident photon will hop to it, but the obtained photons will jump to the other two cavities due to the coherent coupling until the gained photon from the active cavity jumped back to it. Furthermore, at the time $t=\text{arctanh}[\gamma\delta/(\gamma^2-J^2)]/\delta$, the probability for finding the photon in the passive or the central cavities achieve their smallest values simultaneously. On the other hand, the probability for finding a photon in the active cavity will increase monotonously due to the combinational effect of the photonic hopping from the central cavity and the gain from the surrounding environment.

 In Fig.~\ref{bb}(b), we show the results when the single photon is initially excited in the active cavity. In such a situation, the gain effect will take action from the very beginning and the probabilities for finding photons in all of the cavities will undoubtedly increase as the time evolution.

 Comparing the results in Figs.~\ref{bb} (a) and (b), we also observe the unidirectional single-photon transmission in $\mathcal{PT}$ symmetry broken phase. It can be explained from the following two aspects. Firstly, similar to that in $\mathcal{PT}$ symmetric phase,  the time reversal symmetry breaking of the Hamiltonian naturally results in the different transmission behaviors for the photons initially excited in the left and right sides of the system. However, the periodical triangle functions which describe the dynamics of the system in the $\mathcal{PT}$ symmetric phase is replaced by the monotonous hyperbola function in the broken phase. Therefore, the fixed phase difference between $\alpha_s(t)$ and $\xi'_s(t)$ does not hold any longer for $\alpha_b(t)$ and $\xi'_b(t)$. Secondly, the unidirectional transmission also comes from the field localization in the $\mathcal{PT}$ symmetry broken phase. As shown in Sec.~\ref{model}, the long lifetime eigen-state $|E_+\rangle$ has a lager distribution weight in the $1$th cavity, that is, the photon is localized in the active cavity when the system is in the $\mathcal{PT}$ symmetry broken phase. As a result, for the single-photon excited in the active cavity, the overlap between the initial state and $|E_+\rangle$  is much larger than that excited in the passive cavity, and leading to a different transmission behavior.

 \section{conclusion}
 \label{conclusion}
 In conclusion, we have studied the $\mathcal{PT}$ symmetric phase transition by demonstrating
 the spontaneous symmetry breaking in an optical trimmer with balanced loss and gain. In the $\mathcal{PT}$ symmetric phase, all of the eigen values are real and the corresponding eigen states show a balanced distribution in the passive and active cavities. In the $\mathcal{PT}$ symmetry broken phase, the imaginary parts of the eigen values appear and the supermode with long lifetime is characterized by a strong field localization in the active cavity. Comparing with the optical molecule/dimmer system, which was broadly studied recently, the critical coupling strength of the phase transition is much smaller in our trimmer structure and  is therefore easier to be realized experimentally. As a signature of the $\mathcal{PT}$ symmetric phase transition, we subsequently find the dramatically different single-photon transmission behaviors when the system is in the two phases. Our results show that the regular periodical oscillation is replaced by the non-periodical behavior as the system transfers from
 the $\mathcal{PT}$ symmetric phase to $\mathcal{PT}$ symmetry broken phase. Moreover, we find a unidirectional single-photon transmission phenomenon in both phases. We hope our study about the $\mathcal{PT}$ symmetry in optical trimmer will be helpful for the designing of photonic device based on coupled-cavity system.

 {\bf Note added.} Recently, we have learned of recent related work about the $\mathcal{PT}$ symmetry trimmer systems, which cares about the single-photon transmission at the EP~\cite{Jin}.

\begin{acknowledgments}
This work is supported by the National Natural Science Foundation of China (under Grant Nos. 11404021 and 11504241), and the Fundamental Research Funds for the Central Universities (under Grant Nos. 2412016KJ015 and 2412016KJ004).

\end{acknowledgments}

\end{document}